\documentclass[preprint,12pt]{elsarticle}
\usepackage{amssymb}
\usepackage{amsmath}
\usepackage[perpage,symbol]{footmisc}
\usepackage[colorlinks, citecolor=blue]{hyperref}

\journal{Physica A}

\begin{document}

\begin{frontmatter}

\title{Solving Langevin equation with the bicolour rooted tree method}

\author{Jiabin You\footnote[1]{Corresponding author.\\ E-mail address: jiabinyou@gmail.com, zhaoh@xmu.edu.cn}, Hong Zhao}

\address{Department of Physics, Institute of Theoretical Physics and
Astrophysics, Xiamen University, Xiamen 361005, China}

\begin{abstract}
Stochastic differential equations, especially the one called
Langevin equation, play an important role in many fields of modern
science. In this paper, we use the bicolour rooted tree method,
which is based on the stochastic Taylor expansion, to get the
systematic pattern of the high order algorithm for Langevin
equation. We propose a popular test problem, which is related to the
energy relaxation in the double well, to test the validity of our
algorithm and compare our algorithm with other usually used
algorithms in simulations. And we also consider the time-dependent
Langevin equation with the Ornstein-Uhlenbeck noise as our second
example to demonstrate the versatility of our method.
\end{abstract}


\end{frontmatter}

\section{Introduction}

Nature is full of randomness, from nucleus to whole galaxy, from
inorganism to organism and from the domain of science and technology
to the human society \cite{1,2,3,4,5,6,7,8}. Although the mechanisms
of randomness are different from one field to another, the ways to
describe them are similar. The stochastic differential equation
(SDE) is a good approach to describe the randomness. The earliest
work on SDEs was done to describe Brownian motion in Einstein's
famous paper and at the same time by Smoluchowski. Later It\^{o} and
Stratonovich put SDEs on a more solid theoretical foundation. In
1908, a French physicist, Paul Langevin, developed an equation
called the Langevin equation (LE) thereafter, which incorporated a
random force into the Newton equation, to describe the Brownian
motion. Langevin equation is an equation to mechanics using
simplified models and using SDEs to account for omitted degrees of
freedom. There are many branches with rich contents which have been
derived in the last 100 years. For example, the reaction kinetic
dynamics in chemistry \cite{9}, the molecular motor and protein
folding in biology \cite{10,11}, the intracellular and intercellular
calcium signaling, quantum Brownian motion and the stochastic
quantization in physics \cite{12,13,14}, even the stock market
fluctuations and crashes \cite{8} are all related to the Langevin
equation. The Langevin equation plays an important role in modern
science, however only a few of them can be analytically solved, thus
it is necessary to develop a numerical algorithm which incorporates
both the computation efficiency and accuracy.

The general structure of the stochastic differential equation is
\begin{equation}
\dot{x_{i}}=f_{i}(X(t))+g_{i}(X(t))\xi_{i}(t)\ \ \ (i=1,2,...,d),
\end{equation}
where $X(t)=\{x_{1},x_{2},...,x_{d}\}$, $f_{i}(X(t))$s are the
deterministic part of the equations of motion, $g_{i}(X(t))$s are
the diffusion coefficients and $\xi_{i}(t)$s are a set of
independent gaussian random variables with correlation function
\begin{equation}
<\xi_{i}(t)\xi_{j}(t^{'})>=\delta_{ij}\delta(t-t^{'}).
\end{equation}

To get a certain order algorithm for the SDE, we can directly do the
stochastic Taylor expansion of Eq.(1) to our desired accuracy
\cite{15,16}. This method is very explicit and suits for many cases
of the SDEs, however, since this expansion is too laborious to
generate to high orders, we need to find a systematic pattern to
overcome such difficulty. In this paper, we use the bicolour rooted
tree method (BRT) based on the stochastic Taylor expansion to obtain
the high order algorithm for SDEs systematically.

In the field of numerical method for solving ordinary differential
equations, J. C. Butcher develops a rooted tree method which relates
each term in the ordinary Taylor expansion to a rooted tree
\cite{17}. His method can excellently make the laborious ordinary
Taylor expansion systematic in a heuristic way. Then K. Burrage and
P. M. Burrage expand the rooted-tree method to the bicolour rooted
tree method which relates each term in the stochastic Taylor
expansion to a bicolour rooted tree \cite{16} for the sake of
solving SDE. They give an explicit Runge-Kutta method of order 2.5
in their paper for the SDE. In this paper, we further develop their
works, propose a new type of the bicolour rooted tree method, and
apply it to the LE case.

Since the intricacy of the numerical method for SDE, the order of it
is heretofore not great than 2.5 \cite{16,18}. But for some special
kinds of the SDE, for example, the Langevin equation, a high order
algorithm can be acquired. Hershkovitz has developed a fourth order
algorithm for the LE \cite{15}, which is based on the stochastic
Taylor expansion. In this paper, we use the BRT method to improve
the accuracy to order 7 of deterministic part and order 4.5 of
stochastic part (\textit{o}(7,4.5)).

In section 2, we briefly introduce the BRT method and explore the
relation between the terms in the stochastic Taylor expansion and
the bicolour rooted trees. We find that the stochastic Taylor
expansion is just equal to the sum of all the non-isomorphic
bicolour rooted trees. In section 3, due to the structure of LE, we
can use the BRT method to obtain our algorithm \textit{o}(7,4.5) for
the LE. In section 4, we use two examples to verify the validity and
demonstrate the versatility of our algorithm. The first one is the
energy relaxation in the double well. We compare our results with
the previous results obtained by other algorithms and show the
convergence of these different algorithms. The second one we present
an algorithm for the time-dependent Langevin equation with the
Ornstein-Uhlenbeck noise, and our results are readily agreed with
the previous ones.

\section{Bicolour rooted tree method}

To cope with the intricacy of the Taylor expansion of SDE, a method
which is called bicolour rooted tree method (BRT) \cite{16} based on
the rooted tree method \cite{17} developed by J. C. Butcher is
introduced to conveniently do the stochastic Taylor expansion of
SDE.

Let us firstly transform the Eq.(1) into the following equation,
\begin{equation}
dx_{i}=f_{i}(X(t))dt+g_{i}(X(t))\circ dW_{i}(t)\ \ \ (i=1,2,...,d),
\end{equation}
where $W_{i}(t)$ is the Wiener process, and the symbol $\circ$
implies that the SDE considered in this paper is in the Stratonovich
sense, for the Stratonovich integral satisfies the usual rules of
calculus \cite{18}. One can therefore integrate Eq.(3) from 0 to h,
\begin{equation}
x_{i}(h)-x_{i}(0)=\int_{0}^{h}f_{i}(X(s))ds+\int_{0}^{h}g_{i}(X(s))\circ
dW_{i}(s).
\end{equation}
Taylor expansion of the functions gives,
\begin{equation}
f_{i}(X(s))=\sum\limits_{n=0}^{\infty}\frac{1}{n!}(\sum\limits_{m=1}^{d}\delta
x_{m}\frac{\partial}{\partial x_{m}})^{n}f_{i}(X(0)),
\end{equation}
\begin{equation}
g_{i}(X(s))=\sum\limits_{n=0}^{\infty}\frac{1}{n!}(\sum\limits_{m=1}^{d}\delta
x_{m}\frac{\partial}{\partial x_{m}})^{n}g_{i}(X(0)).
\end{equation}
Now taking the last two equations into Eq.(4), one can easily get
\begin{equation}
\begin{split}
\delta x_{i}(h)=x_{i}(h)-x_{i}(0)=&\\
\int_{0}^{h}(f^{i}+f^{i}_{j}\delta
x_{j}(s)+\frac{1}{2!}f^{i}_{jk}&\delta x_{j}(s)\delta
x_{k}(s)+\cdots)ds+\\
\int_{0}^{h}(g^{i}+g^{i}_{j}\delta
x_{j}(s)+\frac{1}{2!}g^{i}_{jk}&\delta x_{j}(s)\delta
x_{k}(s)+\cdots)\circ dW_{i}(s),
\end{split}
\end{equation}
where $f^{i}\equiv f_{i}(X(0)), f^{i}_{jk}\equiv
\frac{\partial}{\partial x_{j}}\frac{\partial}{\partial
x_{k}}f_{i}(X(0))$ and the repeated indices except $i$ (the number
of the equations) imply the Einstein's summation convention
throughout the paper.

Then the terms with 0th derivative in Eq.(7) are,
\begin{equation}
\delta x_{i}^{0}(h)=f^{i}J_{0,h}+g^{i}J_{i,h},
\end{equation}
where $J_{0,h}=\int_{0}^{h}ds$ and $J_{i,h}=\int_{0}^{h}\circ
dW_{i}(s)$, so $\delta x_{i}(h)=\delta x_{i}^{0}(h)+\cdots$,
substituting it for Eq.(7) gives the 1st derivative terms,
\begin{equation}
\delta
x_{i}^{1}(h)=f^{i}_{j}f^{j}J_{00,h}+f^{i}_{j}g^{j}J_{j0,h}+g^{i}_{j}f^{j}J_{0i,h}+g^{i}_{j}g^{j}J_{ji,h},
\end{equation}
where $J_{j_{1}j_{2}\cdots j_{k},t}$ is the Stratonovich multiple
integral \cite{18}, and the integration is with respect to ds if
$j_{l}=0$ or $\circ dW_{i}(s)$ if $j_{l}=i$, for example,
\begin{equation}
J_{012,t}=\int_{0}^{t}\circ dW_{2}(s_{1})\int_{0}^{s_{1}}\circ
dW_{1}(s_{2})\int_{0}^{s_{2}}ds_{3}.
\end{equation}
Replacing Eq.(7) by $\delta x_{i}(h)=\delta x_{i}^{0}(h)+\delta
x_{i}^{1}(h)+\cdots$, one can get the 2nd derivative terms $\delta
x_{i}^{2}(h)$ and performing this procedure recursively will
generate all the derivative terms in principle. However, close
calculation of these derivative terms reveals that the complexity
will increase drastically as the order rises. For this reason, we
adopt the BRT method developed by J. C. Butcher and P. M. Burrage to
express each derivative term systematically and graphically.

We will first introduce some useful notations \cite{16}. Take the
bicolour rooted tree \textit{t} in Fig.1 as an example, The tree has
8 vertices, each vertex can be colored by white node ($\circ$) or
black node ($\bullet$) which is the representative of stochastic
node ($\sigma$) or deterministic node ($\tau$). If
$t_{1},\cdots,t_{m}$ are bicolour rooted trees, then
$[t_{1},\cdots,t_{m}]$ and $\{t_{1},\cdots,t_{m}\}$ are trees in
which $t_{1},\cdots,t_{m}$ are each joined by a single branch to
$\bullet$ or $\circ$, respectively. We can therefore rewrite the
tree \textit{t} in a compact form
$[\sigma,[\{\sigma\},[\tau,\sigma]]]$. To conveniently calculate the
weight of this tree, we define the following terms: the
\textit{degree} of the vertex $C(v)(v\in t)$ in the BRT is
equivalent to the degree of vertex $D(v)$ in the graph theory except
the root 1 with $C(1)=1+D(1)$. S is the \textit{symmetry factor} of
the tree \textit{t}, for example, the trees interchanged the
branches joint to vertex 1 or 3 or 5 are regarded as identical with
tree \textit{t}, therefore the symmetry factor of tree \textit{t} is
$2\times2\times2=8$. Then tracing the stochastic Taylor expansion of
the Eq.(7), we find that the \textit{elementary weight} of the tree,
which is also the coefficient of each term in the expansion, is
\begin{equation}
a(t)=S\prod\limits_{i=1}^{n}\frac{1}{(C(v_{i})-1)!},
\end{equation}
where n is the total number of vertex in the tree and vertex
$v_{i}\in t$. Now we introduce the \textit{elementary derivative}
and \textit{elementary integral} here \cite{16}. An elementary
derivative $F(t)$ can be associated with a BRT such that
\begin{equation}
\begin{split}
&F(\tau)=f, F(\sigma)=g,\\
&F(t)=
\begin{cases}
f^{(m)}[F(t_{1}),\cdots,F(t_{m})], t=[t_{1},\cdots,t_{m}]\\
g^{(m)}[F(t_{1}),\cdots,F(t_{m})], t=\{t_{1},\cdots,t_{m}\},
\end{cases}
\end{split}
\end{equation}
and the elementary integral can be written as
\begin{equation}
\begin{split}
&\theta_{s}(\tau)=J_{0,s}, \theta_{s}(\sigma)=J_{i,s},\\
&\theta_{s}(t)=
\begin{cases}
\int_{0}^{s}du(\prod\limits_{j=1}^{m}\theta_{u}(t_{j})),t=[t_{1},\cdots,t_{m}]\\
\int_{0}^{s}\circ
dW_{i}(u)(\prod\limits_{j=1}^{m}\theta_{u}(t_{j})),t=\{t_{1},\cdots,t_{m}\}.
\end{cases}
\end{split}
\end{equation}
Fig.1(a) illustrates the elementary weight, derivative and integral
graphically.
\begin{figure}
\centerline{
\includegraphics[width=10cm]{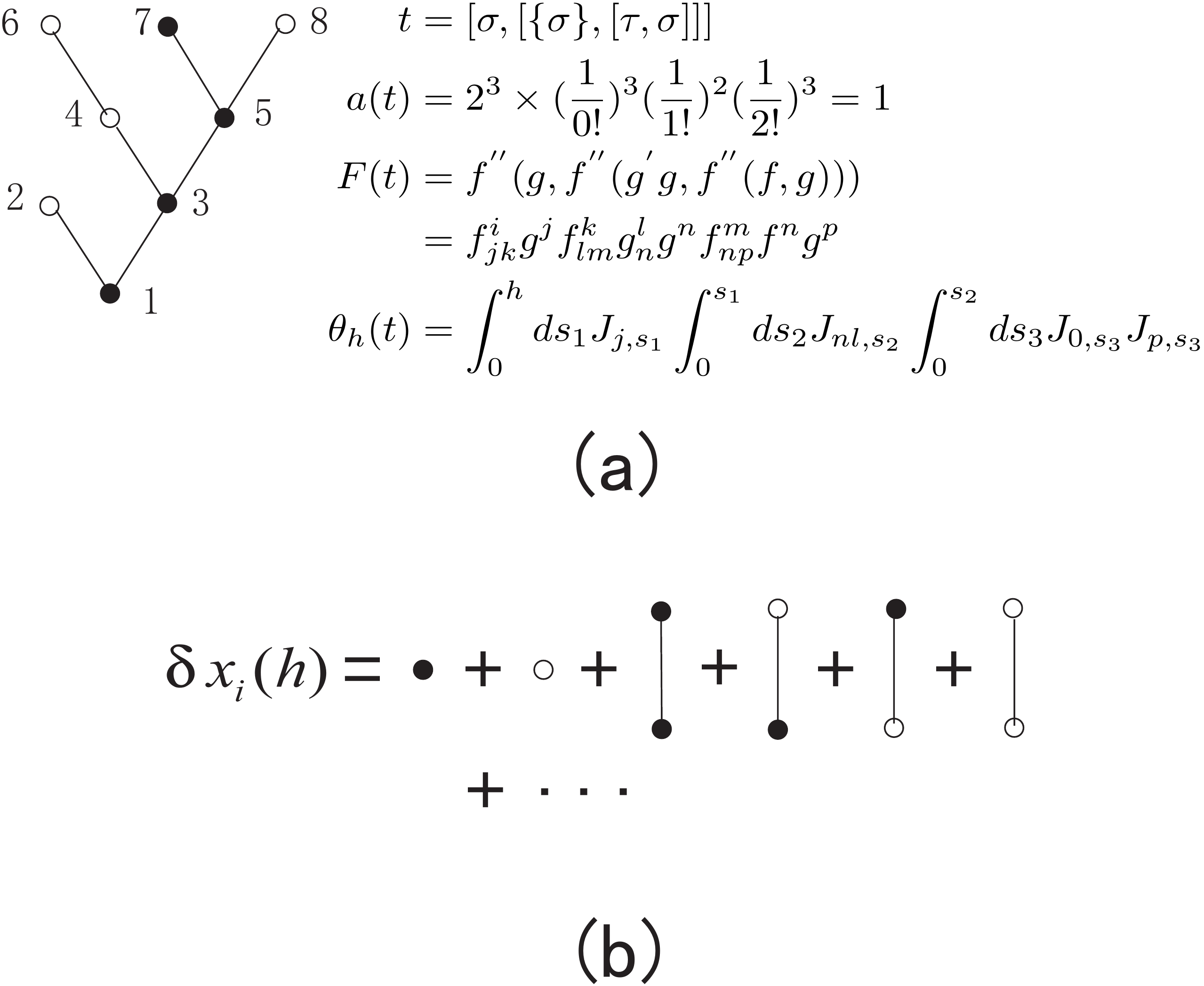}}
\caption{Illustration of the BRT method. Fig.1(a) shows a bicolour
rooted tree t, and its elementary weight $a(t)$, derivative $F(t)$
and integral $\theta_{h}(t)$, respectively. Fig.1(b) shows the 0th
and 1st derivative terms of stochastic Taylor expansion of $\delta
x_{i}(h)$ by the BRT method.}
\end{figure}

Therefore the stochastic Taylor expansion is given by
\begin{equation}
\begin{split}
\delta x_{i}(h)=&\sum\limits_{t\in T}a(t)F(t)\theta_{h}(t)\\
=&\textit{sum of all the non-isomorphic}\\
&\textit{bicolour rooted trees},\\
\end{split}
\end{equation}
where T is the set of non-isomorphic bicolour rooted trees. Fig.1(b)
illustrates how to use this formula to express $\delta
x_{i}(h)=\delta x_{i}^{0}(h)+\delta x_{i}^{1}(h)+\cdots$.

\section{Algorithm}

Due to the complexity of stochastic Taylor expansion, we only
consider Langevin equation (LE) which plays an important part in the
fields involving randomness in this paper. An N dimensional set of
coupled LEs has the form
\begin{equation}
\ddot{q_{i}}=-\frac{\partial V(\textbf{q}(t))}{\partial
q_{i}}-\gamma_{i}\dot{q_{i}}+\xi_{i}(t),
\end{equation}
where $V(\textbf{q}(t))$ is the external potential,
$\gamma_{i}(i=1,\cdots,N)$ is a set of friction parameters,
$\xi_{i}(t)$ is random noise with zero mean, the correlation
relation is
\begin{equation}
<\xi_{i}(t)\xi_{j}(t^{'})>=2\gamma_{i}T\delta_{ij}\delta(t-t^{'}),
\end{equation}
and the Hamilton canonical equations are
\begin{equation}
\begin{split}
&\dot{q_{i}}=p_{i},\\
&\dot{p_{i}}=-\frac{\partial V(\textbf{q}(t))}{\partial
q_{i}}-\gamma_{i}p_{i}+\xi_{i}(t).
\end{split}
\end{equation}
The form of Eq.(17) where only every second equation has a noise
term with constant diffusion coefficient, as well as the potential
$V(\textbf{q}(t))$ is unrelated to $\textbf{p}(t)$, makes it
possible to sharply decrease the complexity of Eq.(14) so as to
obtain a high order algorithm for LE.

For the Eq.(17), we can translate it into the form
\begin{equation}
\begin{split}
&\dot{x_{i}}=f_{i}(X(t))+g_{i}\xi_{i}(t)\ \ \ (i=1,2,...,2N)\\
&<\xi_{i}(t)\xi_{j}(t^{'})>=\delta_{ij}\delta(t-t^{'}),
\end{split}
\end{equation}
where $f_{i}(X(t))$ is equal to $x_{i+1}$ for odd \textit{i} and to
$-\partial V(\textbf{X}(t))/\partial x_{i-1}-\gamma_{i}x_{i}$ for
even \textit{i}, $\textbf{X}(t)=\{x_{1},x_{3},...,x_{2N-1}\}$,
$g_{i}$ is a set of constants with $g_{i}=0$ if \textit{i} is odd
number. Because of the property of $V(\textbf{q}(t))$, one can find
that $f_{jk}^{i}\neq0$ only if \textit{j} and \textit{k} are both
odd numbers. From above properties, a key property for the
simplification of the stochastic Taylor expansion, that is, $\cdots
f_{jk}^{i}g_{j}\cdots=0$, can be found. We can rewrite it in the
compact form as follow:
\begin{equation}
\cdots,[\cdots,\sigma,\cdots],\cdots=0,
\end{equation}
so if a bicolour rooted tree has this structure, it should have no
contribution to the stochastic Taylor expansion.

From the analysis above, one can obtain a general method for solving
the Langevin equation numerically. If we want to get a numerical
method to the order \textit{o(m,n)}, we should:

(a)For the deterministic part:

Solve it by the standard Runge-Kutta method of order m.

(b)For the stochastic part:

(i)Write down all the non-isomorphic bicolour rooted trees that can
avoid the structure (19);

(ii)Attach each vertex with white or black so as to make the tree
have order n.

(c)Add up the results of (a) and (b).

Using these three criteria, all the terms up to order
\textit{o}(7,4.5) are,
\begin{equation}
\begin{split}
\delta x_{idet}(h)=&\sum\limits_{j=0}^{6}RK(j),\\
\delta x_{iran}(h)=&\sigma+[\sigma]+[[\sigma]]+[[[\sigma]]]+[\tau,[\sigma]]\\
&+[[[[\sigma]]]]+[\tau,[[\sigma]]]+[[\tau],[\sigma]]\\
&+[\tau,\tau,[\sigma]]+[[\tau,[\sigma]]]+[[\sigma],[\sigma]],\\
\delta x_{i}(h)=&\delta x_{idet}(h)+\delta x_{iran}(h).\\
\end{split}
\end{equation}
where the $RK(j),(j=0,\cdots,6)$ are the BRTs with $j+1$
deterministic nodes only which are identical to the terms of
standard Runge-Kutta method for ODEs. These terms compose the
deterministic part of our algorithm, and we can use Runge-Kutta
method to solve the deterministic part numerically \cite{17,19}.

Then we try to find a way to calculate the stochastic part in
Eq.(20). We here introduce a method to approximate the elementary
integral which is developed by P. E. Kloeden and E. Platen
\cite{18}. They showed that if
$\textbf{W}(t)=(W_{1}(t),\cdots,W_{N}(t))$ is an N-dimensional
Wiener process on the time interval $[0,h]$, the componentwise
Fourier expansion of the Brownian bridge process
$\textbf{W}(t)-\frac{t}{h}\textbf{W}(h)$ is
\begin{equation}
W_{i}(t)-\frac{t}{h}W_{i}(h)=\frac{1}{2}a_{i0}+\sum\limits_{j=1}^{\infty}(a_{ij}\cos(\frac{2j\pi
t}{h})+b_{ij}\sin(\frac{2j\pi t}{h})),
\end{equation}
where $a_{ij},b_{ij}$ are $N(0,h/2\pi^{2}j^{2})$ distributed and
pairwise independent, then setting $t=0$ in the equation (21) gives
$a_{i0}=-2\sum\limits _{j=1}^{\infty}a_{ij}\equiv a_{i}^{0}$.

Now, we begin to calculate the stochastic part of Eq.(20). Firstly,
let us set the $\textbf{W}(0)=0$, then use equation (21), we can
easily find that
\begin{equation}
\begin{split}
\sigma&=a(\sigma)F(\sigma)\theta_{h}(\sigma)=g^{i}J_{i,h}=g^{i}W_{i}(h)\equiv g^{i}\omega_{i}^{1},\\
[\sigma]&=f_{j}^{i}g^{j}J_{j0,h}=f_{j}^{i}g^{j}\int_{0}^{h}ds_{1}\int_{0}^{s_{1}}\circ
dW_{j}(s_{2})\\
&=f_{j}^{i}g^{j}\int_{0}^{h}W_{j}(s_{1})ds_{1}\\
&=hf_{j}^{i}g^{j}(\frac{W_{j}(h)}{2}+\frac{a_{j}^{0}}{2})\\
&\equiv hf_{j}^{i}g^{j}\omega_{j}^{2},\\
\end{split}
\end{equation}
where $W_{i}(h)\equiv W_{i}$ is a set of independent Gaussian random
variables sampled from $N(0,h)$. Similarly calculation gives
\begin{equation}
\begin{split}
[[\sigma]]&=h^{2}f_{j}^{i}f_{k}^{j}g^{k}\omega_{k}^{3}\\
[[[\sigma]]]&=h^{3}f_{j}^{i}f_{k}^{j}f_{l}^{k}g^{l}\omega_{l}^{4}\\
[\tau,[\sigma]]&=h^{3}f_{jk}^{i}f^{j}f_{l}^{k}g^{l}\omega_{l}^{5}\\
[[[[\sigma]]]]&=h^{4}f_{j}^{i}f_{k}^{j}f_{l}^{k}f_{m}^{l}g^{m}\omega_{m}^{6}\\
[\tau,[[\sigma]]]&=h^{4}f_{jk}^{i}f^{j}f_{l}^{k}f_{m}^{l}g^{m}\omega_{m}^{7}\\
[[\tau],[\sigma]]&=h^{4}f_{jk}^{i}f_{l}^{j}f^{l}f_{m}^{k}g^{m}\omega_{m}^{8}\\
[\tau,\tau,[\sigma]]&=\frac{h^{4}}{2}f_{jkl}^{i}f^{j}f^{k}f_{m}^{l}g^{m}\omega_{m}^{9}\\
[[\tau,[\sigma]]]&=h^{4}f_{j}^{i}f_{kl}^{j}f^{k}f_{m}^{l}g^{m}\omega_{m}^{10}\\
[[\sigma],[\sigma]]&=\frac{h^{3}}{2}f_{jk}^{i}f_{l}^{j}g^{l}f_{m}^{k}g^{m}\Omega_{lm}^{1}\\
\end{split}
\end{equation}
where
\begin{equation}
\begin{split}
\omega_{i}^{3}&=\frac{W_{i}}{6}+\frac{a_{i}^{0}}{4}+\frac{b_{i}^{1}}{2}\\
\omega_{i}^{4}&=\frac{W_{i}}{24}+\frac{a_{i}^{0}}{12}+\frac{a_{i}^{1}}{4}+\frac{b_{i}^{1}}{4}\\
\omega_{i}^{5}&=\frac{W_{i}}{8}+\frac{a_{i}^{0}}{6}-\frac{a_{i}^{1}}{4}+\frac{b_{i}^{1}}{4}\\
\omega_{i}^{6}&=\frac{W_{i}}{120}+\frac{a_{i}^{0}}{48}+\frac{a_{i}^{1}}{8}+\frac{b_{i}^{1}}{12}-\frac{b_{i}^{2}}{8}\\
\omega_{i}^{7}&=\frac{W_{i}}{30}+\frac{a_{i}^{0}}{16}+\frac{a_{i}^{1}}{8}+\frac{b_{i}^{1}}{6}+\frac{b_{i}^{2}}{8}\\
\omega_{i}^{8}&=\frac{W_{i}}{20}+\frac{a_{i}^{0}}{16}-\frac{a_{i}^{1}}{8}+\frac{b_{i}^{1}}{12}-\frac{b_{i}^{2}}{8}\\
\omega_{i}^{9}&=\frac{W_{i}}{10}+\frac{a_{i}^{0}}{8}-\frac{a_{i}^{1}}{4}+\frac{b_{i}^{1}}{6}-\frac{b_{i}^{2}}{4}\\
\omega_{i}^{10}&=\frac{W_{i}}{40}+\frac{a_{i}^{0}}{24}+\frac{b_{i}^{1}}{12}+\frac{b_{i}^{2}}{4}\\
a_{i}^{0}&=-2\sum\limits_{j=1}^{\infty}a_{ij},\ b_{i}^{1}=\sum\limits_{j=1}^{\infty}\frac{1}{\pi j}b_{ij},\\
a_{i}^{1}&=\sum\limits_{j=1}^{\infty}\frac{1}{(\pi j)^{2}}a_{ij},\ b_{i}^{2}=\sum\limits_{j=1}^{\infty}\frac{1}{(\pi j)^{3}}b_{ij},\\
\end{split}
\end{equation}
and the first non-Gaussian random variable is
\begin{equation}
\begin{split}
\Omega_{ij}^{1}&=\omega_{i}^{1}\omega_{j}^{9}+a_{i}^{0}\omega_{j}^{5}-\frac{\omega_{i}^{1}\omega_{j}^{1}}{20}-\frac{a_{i}^{0}a_{j}^{0}}{12}-\frac{a_{i}^{0}\omega_{j}^{1}}{8}+\frac{b_{i}^{1}b_{j}^{1}}{4}+\frac{c_{ij}}{8},\\
c_{ij}&=\sum\limits_{k=1}^{\infty}\frac{1}{(\pi
k)^{2}}(a_{ik}a_{jk}+b_{ik}b_{jk}).\\
\end{split}
\end{equation}
We can find that there are only 5 independent variables among
$\omega_{i}^{j},(j=1,\cdots,10)$. Let's choose
$\omega_{i}^{1},\omega_{i}^{2},\omega_{i}^{3},\omega_{i}^{4},\omega_{i}^{6}$
as the independent variables, then
\begin{equation}
\begin{split}
\omega_{i}^{5}&=\omega_{i}^{3}-\omega_{i}^{4}\\
\omega_{i}^{7}&=\omega_{i}^{4}-\omega_{i}^{6}\\
\omega_{i}^{8}&=\frac{1}{2}\omega_{i}^{3}-\omega_{i}^{4}+\omega_{i}^{6}\\
\omega_{i}^{9}&=\omega_{i}^{3}-2\omega_{i}^{4}+2\omega_{i}^{6}\\
\omega_{i}^{10}&=\omega_{i}^{4}-2\omega_{i}^{6}
\end{split}
\end{equation}

Now, the last procedure we should do is to determine the five
Gaussian random variables
$W_{i},a_{i}^{0},a_{i}^{1},b_{i}^{1},b_{i}^{2}$ and the non-Gaussian
random variable $c_{ij}$. We truncate $c_{ij}$ to the first term,
that is, $c_{ij}\approx(a_{i1}a_{j1}+b_{i1}b_{j1})/\pi^{2}$. Since
$W_{i},a_{ij},b_{ij}$ are independent and $W_{i}\sim N(0,h),\
a_{ij}\sim N(0,h/2\pi^{2}j^{2}),\ b_{ij}\sim N(0,h/2\pi^{2}j^{2})$,
we can see that
\begin{equation}
\begin{split}
a_{i}^{0}\sim& N(0,\frac{h}{3}),\ b_{i}^{1}\sim N(0,\frac{h}{180}),\\
a_{i}^{1}\sim& N(0,\frac{h}{1890}),\ b_{i}^{2}\sim N(0,\frac{h}{18900}),\\
<a_{i}^{0}&a_{i}^{1}>=-\frac{h}{90},\ <a_{i}^{0}a_{i1}>=-\frac{h}{\pi^{2}},\ <a_{i}^{1}a_{i1}>=\frac{h}{2\pi^{4}},\\
<b_{i}^{1}&b_{i}^{2}>=\frac{h}{1890},\ <b_{i}^{1}b_{i1}>=\frac{h}{2\pi^{3}},\ <b_{i}^{2}b_{i1}>=\frac{h}{2\pi^{5}}.\\
\end{split}
\end{equation}

Let
$\phi_{i}^{1},\phi_{i}^{2},\phi_{i}^{3},\phi_{i}^{4},\phi_{i}^{5},\phi_{i}^{6},\phi_{i}^{7}$
to be seven independent standard Gaussian random variables, then use
Eq.(27), we can get
\begin{equation}
\begin{split}
W_{i}&=\sqrt{h}\phi_{i}^{1}\\
a_{i}^{0}&=\sqrt{\frac{h}{3}}\phi_{i}^{2}\\
a_{i}^{1}&=\frac{\sqrt{h}}{30}(-\frac{\phi_{i}^{2}}{\sqrt{3}}+\frac{\phi_{i}^{3}}{\sqrt{7}})\\
b_{i}^{1}&=\sqrt{\frac{h}{180}}\phi_{i}^{4}\\
b_{i}^{2}&=\frac{\sqrt{h}}{63}(\frac{\phi_{i}^{4}}{\sqrt{5}}+\frac{\phi_{i}^{5}}{10})\\
a_{i1}&=\sqrt{h}(-0.175493\phi_{i}^{2}+0.139348\phi_{i}^{3}+0.0210906\phi_{i}^{6})\\
b_{i1}&=\sqrt{h}(0.21635\phi_{i}^{4}+0.0617995\phi_{i}^{5}+0.00584342\phi_{i}^{7})\\
\end{split}
\end{equation}

The BRT method gives an algorithm for the Langevin equation so long
as we determine the deterministic and the stochastic part of Eq.(20)
respectively and add up each other. The deterministic part can be
solved by the standard Runge-Kutta algorithm \cite{19}, and the
stochastic part can be solved by the Eqs.(22)-(23).

\section{Numerical simulations}

\subsection{Energy relaxation in double well}

To verify the validity of our algorithm, the Kramers equation will
be considered as the severe test for our algorithm. The form is as
follow:
\begin{equation}
\begin{split}
\dot{q}(t)&=p(t),\\
\dot{p}(t)&=-V^{'}(q(t))-\gamma p(t)+\xi(t),\\
\end{split}
\end{equation}
and $\xi(t)$ is the Gaussian random force obeying the fluctuation
dissipation theorem
\begin{equation}
<\xi(t)\xi(t^{'})>=2\gamma T.
\end{equation}
Our method implies that the algorithm for Eqs.(29)-(30) is,
\begin{equation}
\begin{split}
q(t+h)=&q_{det}(q(t),p(t),h)+q_{ran}(q(t),p(t),h),\\
p(t+h)=&p_{det}(q(t),p(t),h)+p_{ran}(q(t),p(t),h),\\
\end{split}
\end{equation}
where $q_{det}(q(t),p(t),h)$ and $p_{det}(q(t),p(t),h)$ are the
results of evolving the equations in the period \textit{0-h} by the
seventh-order Runge-Kutta algorithm \cite{19} which is used in the
ODE, and the stochastic part of Eq.(31) is,
\begin{equation}
\begin{split}
q_{ran}(q(t),p(t),h)=&\sqrt{2\gamma T}[(h\omega_{2}^{2}-h^{2}\gamma\omega_{2}^{3}+h^{3}\gamma^{2}\omega_{2}^{4}-h^{4}\gamma^{3}\omega_{2}^{6})\\
+&(-h^{3}\omega_{2}^{4}+2h^{4}\gamma\omega_{2}^{6})V^{''}-h^{4}p(t)\omega_{2}^{10}V^{'''}],\\
p_{ran}(q(t),p(t),h)=&\sqrt{2\gamma T}[(\omega_{2}^{1}-h\gamma\omega_{2}^{2}+h^{2}\gamma^{2}\omega_{2}^{3}-h^{3}\gamma^{3}\omega_{2}^{4}\\
+&h^{4}\gamma^{4}\omega_{2}^{6})+(-h^{2}\omega_{2}^{3}+2h^{3}\gamma\omega_{2}^{4}-3h^{4}\gamma^{2}\omega_{2}^{6})V^{''}\\
+&(-h^{3}p(t)\omega_{2}^{5}+h^{4}\gamma p(t)\omega_{2}^{7}+h^{4}\gamma p(t)\omega_{2}^{8}\\
+&h^{4}\gamma
p(t)\omega_{2}^{10})V^{'''}+h^{4}V^{''}V^{''}\omega_{2}^{6}+h^{4}V^{'}V^{'''}\omega_{2}^{8}\\
-&\frac{h^{4}}{2}V^{''''}p^{2}(t)\omega_{2}^{9}]-h^{3}\gamma
TV^{'''}\Omega_{22}^{1},
\end{split}
\end{equation}
where $\omega_{2}^{i} (i=1,\cdots,10)$ and $\Omega_{22}^{1}$ have
been defined in the previous section.

The double well potential in this example is,
\begin{equation}
V(q)=q^{4}-2q^{2}.
\end{equation}
It has two minima located at $q=\pm1$ and a potential barrier with
the height $\Delta V=1$ between the two wells. The friction
coefficient $\gamma$ is set to 1. The initial condition is chosen on
the top of the barrier. The average is taken over 5000 realizations
of the Gaussian random force during the trajectory. Fig.2 shows the
result which is compared with the Euler method and the Heun method
\cite{20}. We perform these three methods at T=0.05 and T=0.2
respectively. We find that the results of these three different
methods are almost agreed. Nevertheless, the step size of our
method, Heun method and Euler method are 0.1, 0.001 and 0.0001,
respectively. The Kramers equation has been simulated extensively by
many authors (Ref. \cite{15} and the references therein). As for the
convergence, we compare our algorithm with previous algorithms here.
Fig.(3) shows the convergence of the three algorithms for solving
the Kramers equation. It is evident that our algorithm diverges
slowly than the other algorithms as the step size increases.

\begin{figure}
\centerline{
\includegraphics[width=10cm]{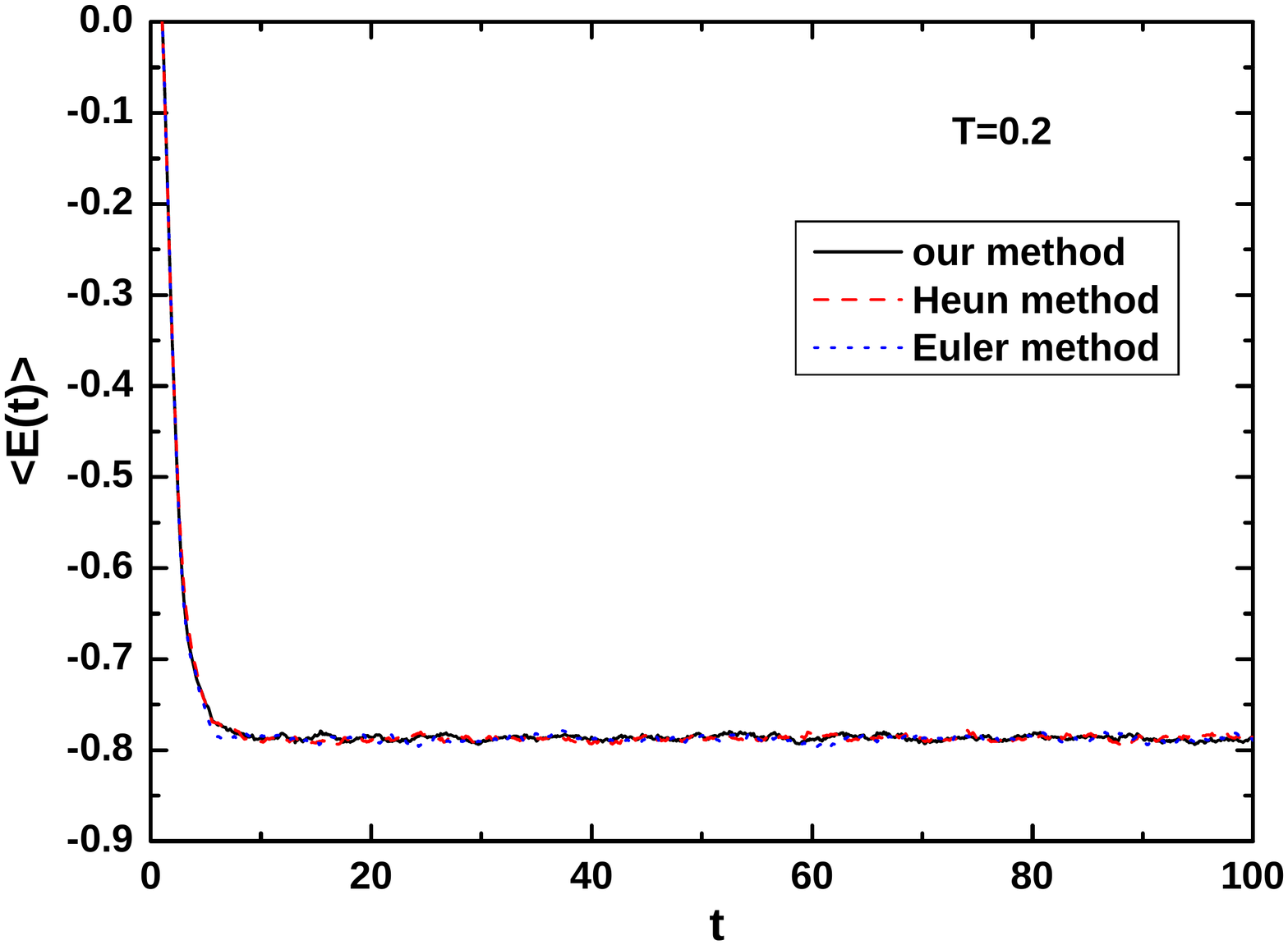}}
\centerline{
\includegraphics[width=10cm]{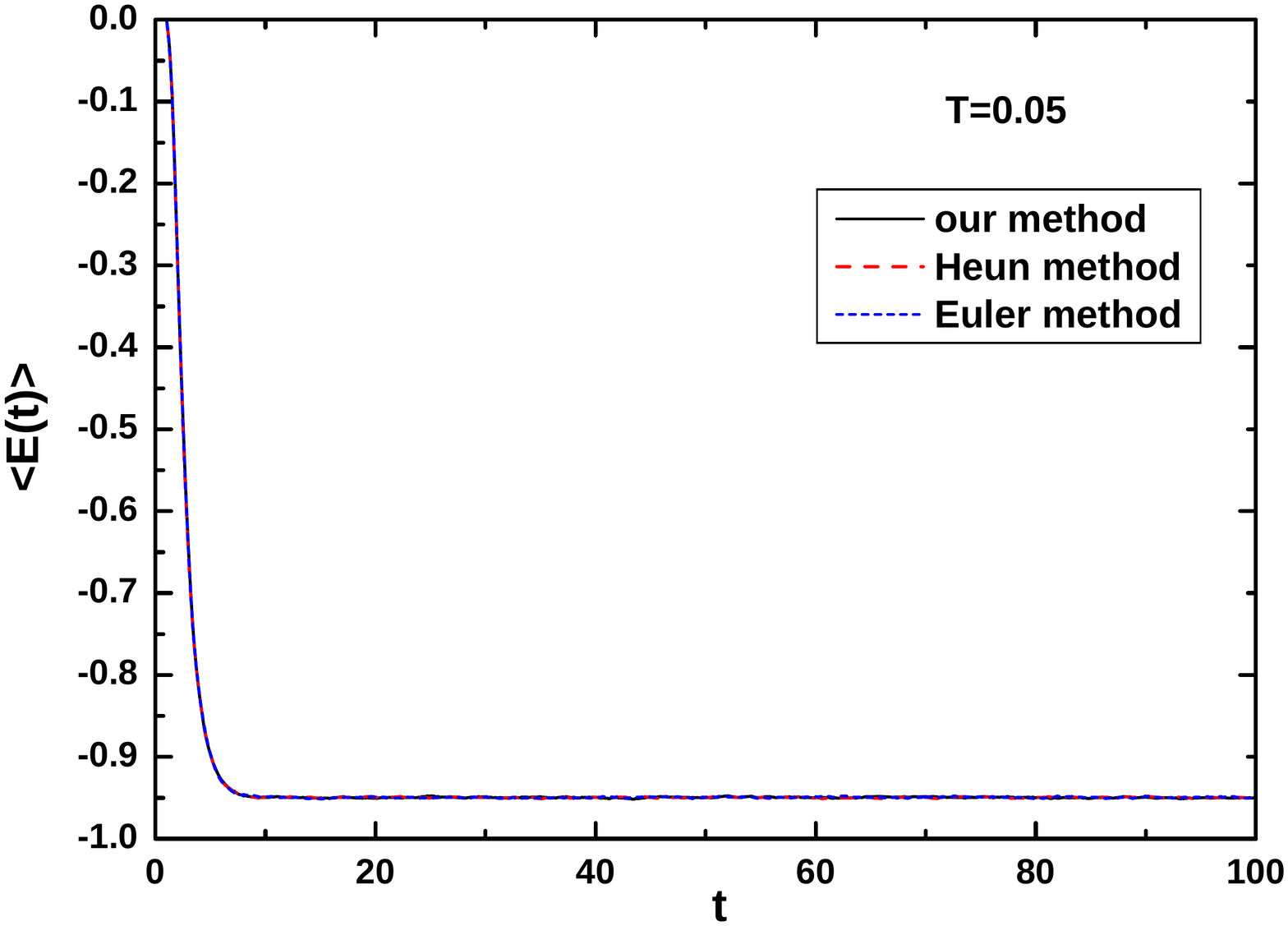}}
\caption{Comparison of different numerical methods for solving the
energy relaxation in the double well potential. The friction
coefficient $\gamma$ is equal to 1. The average is taken over 5000
realizations for all the algorithms. The solid line is our method
with step size 0.1, the dash line is the Heun method with step size
0.001, and the short dash line is the Euler method with the step
size 0.0001. Panel a shows the relaxation at high temperature T=0.2
and panel b shows the relaxation at low temperature T=0.05,
respectively.}
\end{figure}

\begin{figure}
\centerline{
\includegraphics[width=10cm]{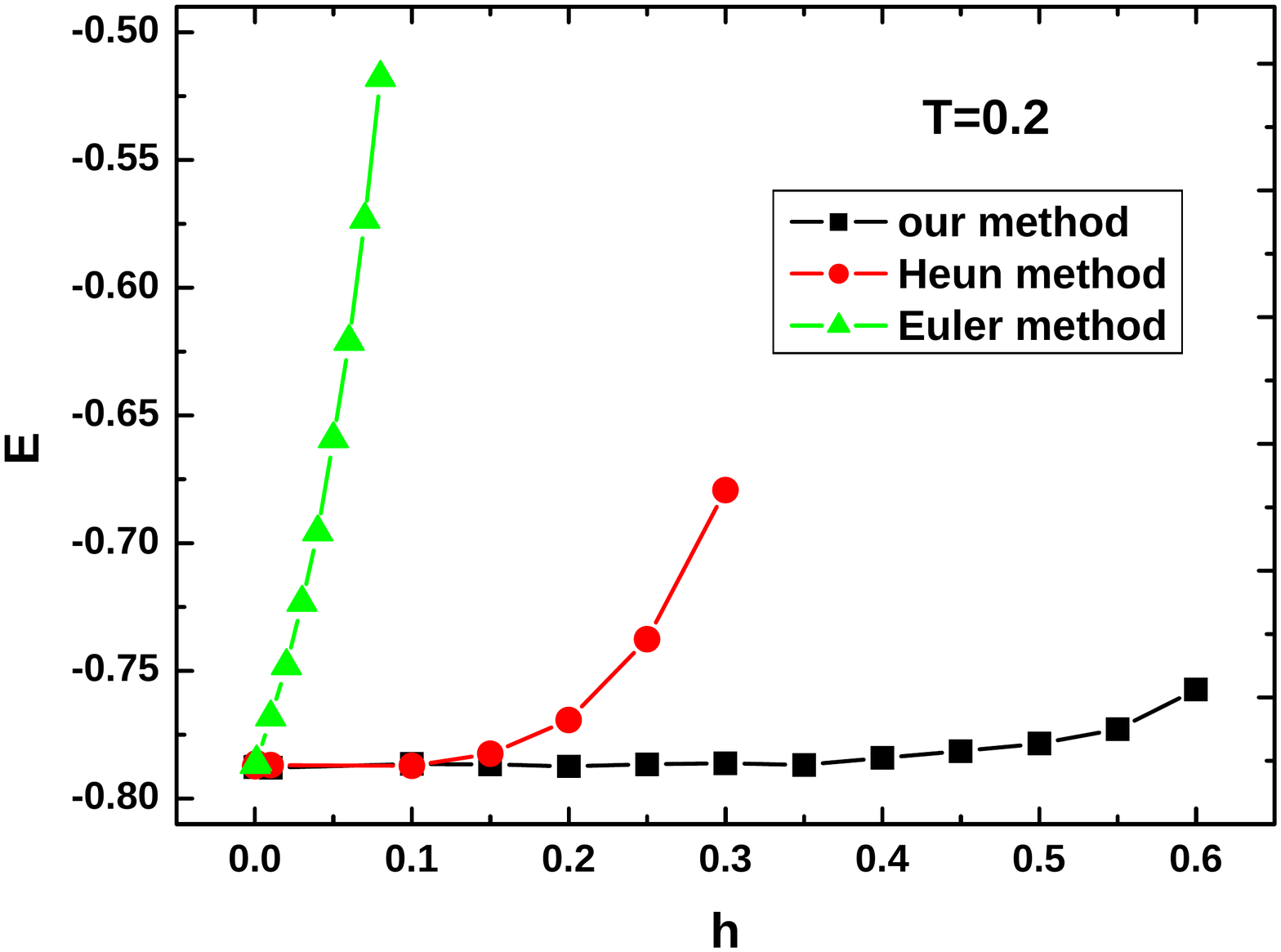}}
\centerline{
\includegraphics[width=10cm]{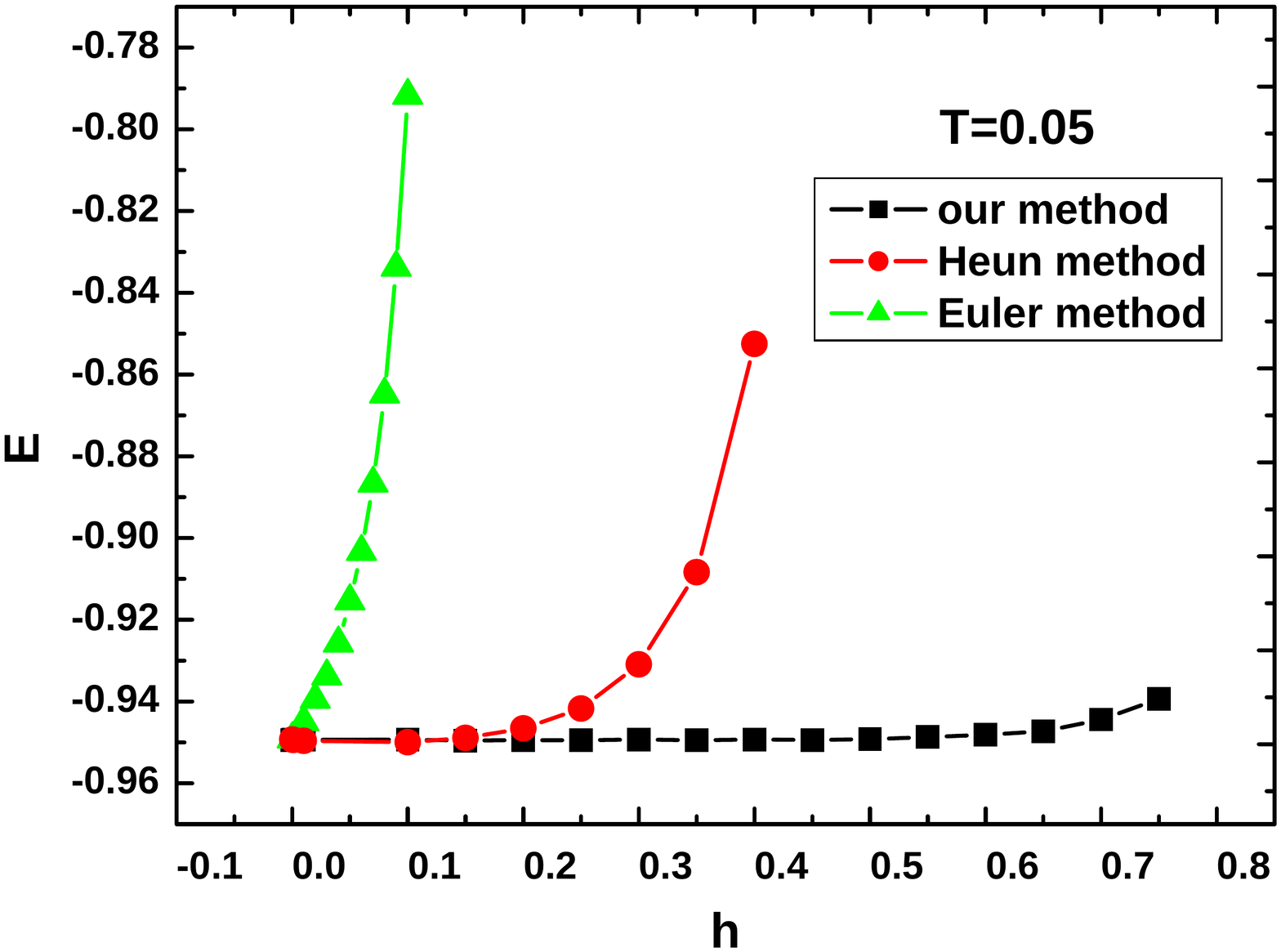}}
\caption{The convergence of different numerical methods for solving
the Kramers equation. The friction coefficient $\gamma$ is equal to
1. E is the average energy at a certain temperature and h is the
step size. Panel a shows the convergence at high temperature T=0.2
and panel b shows the convergence at low temperature T=0.05,
respectively.}
\end{figure}

\subsection{Stochastic resonance}

Stochastic resonance, which is originally developed to explain the
ice ages \cite{21,22}, has spread well beyond physics and left its
fingerprints in many other research areas \cite{23,24}, such as
complex networks \cite{25}, biological systems \cite{26},
neuroscience \cite{27,28} and quantum computing \cite{29}. The
governing equations in these very different fields are essentially
Langevin equation or its generalizations. We present an example of
stochastic resonance in neuroscience to demonstrate our algorithm in
the case of time-dependent Langevin equation with the
Ornstein-Uhlenbeck noise \cite{23}. An enlightening model in the
neuronal dynamical systems is the noise-driven bistable system whose
equations can be described as follow:
\begin{equation}
\begin{split}
&\ddot{x}+\gamma\dot{x}=-V^{'}(x)+A\cos(\omega t)+\varepsilon(t),\\
&<\varepsilon(t)\varepsilon(t^{'})>=D\lambda\exp(-\lambda|t-t^{'}|),\\
\end{split}
\end{equation}
where $\varepsilon(t)$ is the Ornstein-Uhlenbeck noise with
intensity D and the inverse of characteristic time $\lambda$, and
the system is driven by an external periodic force with amplitude A
and frequency $\omega$.

To use our algorithm to solve Eq.(34) numerically, we should first
transform it into,
\begin{equation}
\begin{split}
&\dot{y}=1,\\
&\dot{x}=p,\\
&\dot{p}=-\gamma p-V^{'}(x)+A\cos(\omega y)+\varepsilon,\\
&\dot{\varepsilon}=-\lambda\varepsilon+\lambda\sqrt{2D}\xi(t),\\
&<\xi(t)\xi(t^{'})>=\delta(t-t^{'}).\\
\end{split}
\end{equation}
Let $y\rightarrow x_{1},x\rightarrow x_{2},p\rightarrow
x_{3},\varepsilon\rightarrow x_{4}$, we can further simplify Eq.(35)
into a compact form,
\begin{equation}
\begin{split}
&\dot{x_{i}}=f_{i}(X(t))+g_{i}\xi_{i}(t)\ \ \ (i=1,2,3,4),\\
&<\xi(t)\xi(t^{'})>=\delta(t-t^{'}),\\
\end{split}
\end{equation}
with
\begin{equation}
\begin{split}
&f_{1}(X(t))=1,f_{2}(X(t))=x_{3},f_{4}(X(t))=-\lambda x_{4},\\
&f_{3}(X(t))=-\gamma x_{3}-V^{'}(x_{2})+A\cos(\omega x_{1})+x_{4},\\
&g_{1}=g_{2}=g_{3}=0,\ g_{4}=\lambda\sqrt{2D},\\
\end{split}
\end{equation}
then one can easily find that property (19) is held again.

Accordingly, the numerical method of Eqs.(36)-(37) can be written as
follow:
\begin{equation}
\begin{split}
x_{1}(t+h)=&x_{1det}(X(t),h)+x_{1ran}(X(t),h),\\
x_{2}(t+h)=&x_{2det}(X(t),h)+x_{2ran}(X(t),h),\\
x_{3}(t+h)=&x_{3det}(X(t),h)+x_{3ran}(X(t),h),\\
x_{4}(t+h)=&x_{4det}(X(t),h)+x_{4ran}(X(t),h),\\
\end{split}
\end{equation}
where the deterministic part of Eq.(38) accords with the Ronge-Kutta
algorithm for the ODEs, and the stochastic part of Eq.(38) is
\begin{equation}
\begin{split}
x_{1ran}&(X(t),h)=0,\\
x_{2ran}&(X(t),h)=\lambda\sqrt{2D}[(h^{2}\omega_{4}^{3}-h^{3}(\gamma+\lambda)\omega_{4}^{4}\\
&+h^{4}(\gamma^{2}+\gamma\lambda+\lambda^{2})\omega_{4}^{6})-h^{4}\omega_{4}^{6}V^{''}],\\
x_{3ran}&(X(t),h)=\lambda\sqrt{2D}[(h\omega_{4}^{2}-h^{2}(\gamma+\lambda)\omega_{4}^{3}\\
&+h^{3}(\gamma^{2}+\gamma\lambda+\lambda^{2})\omega_{4}^{4}-h^{4}(\gamma^{3}+\gamma^{2}\lambda\\
&+\gamma\lambda^{2}+\lambda^{3})\omega_{4}^{6})+(-h^{3}\omega_{4}^{4}+h^{4}(2\gamma+\lambda)\omega_{4}^{6})V^{''}\\
&-h^{4}x_{3}(t)\omega_{4}^{7}V^{'''}],\\
x_{4ran}&(X(t),h)=\lambda\sqrt{2D}[\omega_{4}^{1}-h\lambda\omega_{4}^{2}+h^{2}\lambda^{2}\omega_{4}^{3}\\
&-h^{3}\lambda^{3}\omega_{4}^{4}+h^{4}\lambda^{4}\omega_{4}^{6}].\\
\end{split}
\end{equation}

The double well potential in this example is,
\begin{equation}
V(x)=-\frac{1}{2}x^{2}+\frac{1}{4}x^{4},
\end{equation}
and the periodic driven force's amplitude A and frequency $\omega$
are 0.03 and 0.01 respectively. We consider the relation between the
amplitude of output of the system $<x>$ and the noise intensity D.
The average is taken over $5\times10^{5}$ realizations and the Heun
method is used as a comparison. We then compare our results with the
model mentioned in \cite{23}:
\begin{equation}
\begin{split}
&\dot{x}=-V^{'}(x)+A\cos(\omega t)+\xi(t),\\
&<\xi(t)\xi(t^{'})>=2D\delta(t-t^{'}).\\
\end{split}
\end{equation}
The theoretical result of $<x>$ in this model is
$<x>=\frac{A}{D}\frac{2r_{k}}{\sqrt{4r_{k}^{2}+\omega^{2}}}$, where
$r_{k}=\frac{1}{\sqrt{2}\pi}\exp{(-\frac{\Delta V}{D})}$.

\begin{figure}
\centerline{
\includegraphics[width=10cm]{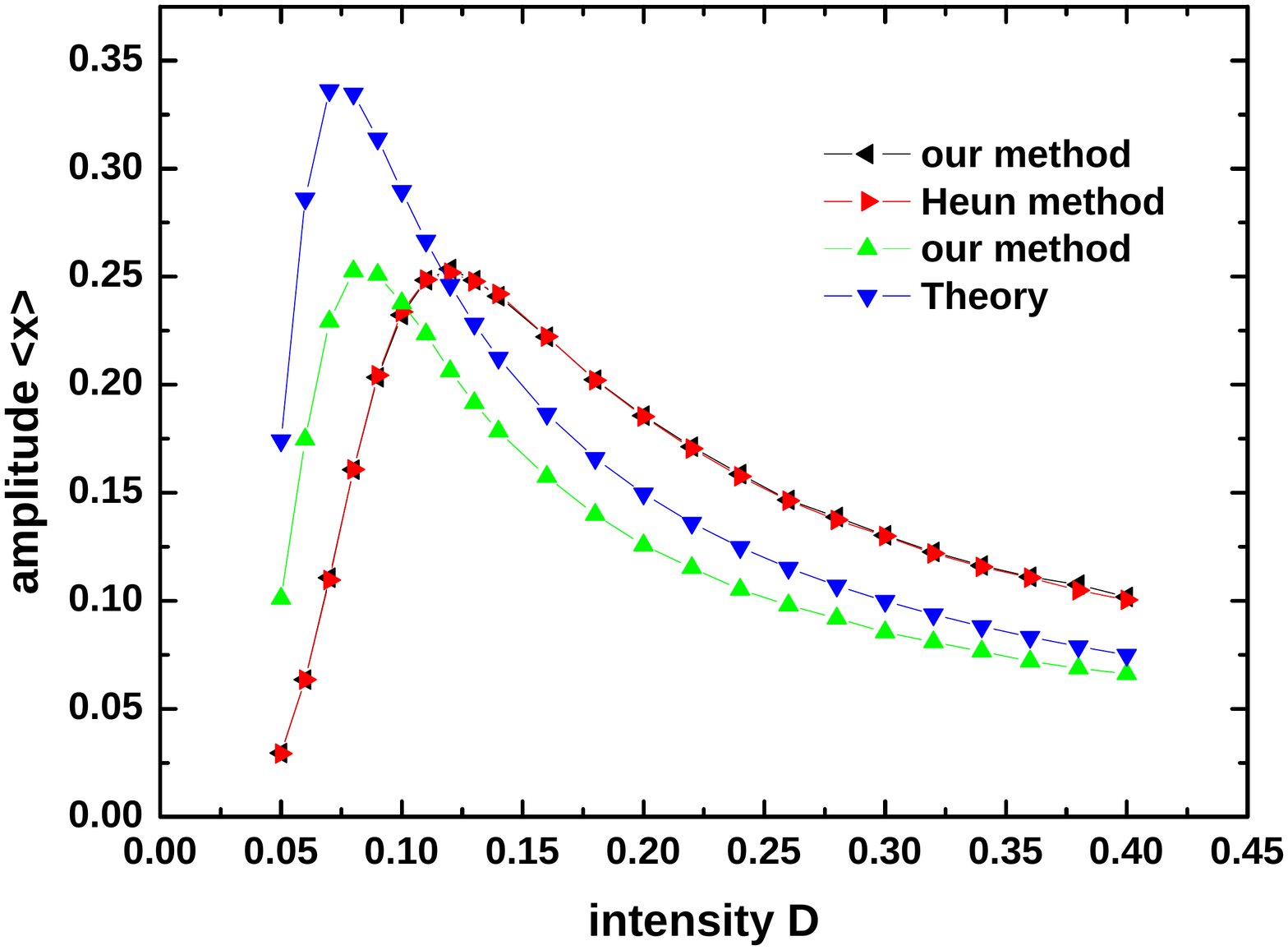}}
\caption{Comparison of different numerical methods for the
stochastic resonance in the noise-driven bistable system. The
periodic driven force's amplitude A and frequency $\omega$ are 0.03
and 0.01, respectively. The friction coefficient $\gamma$ is equal
to 1. The amplitude of output is averaged over $5\times10^{5}$
trajectories. The black line is our method with step size 0.1 and
the red line is the Heun method with step size 0.01. Both the two
lines have the characteristic time 1. The green line is our method
with step size 0.1 and characteristic time 0.1. The blue line is the
theoretical result of Eq.(41).}
\end{figure}

Fig.(4) shows the results of our simulations. The black line and the
red line are the simulations of our method with step size 0.1 and
the Heun method with step size 0.01 respectively, with the
parameters $\gamma$ and $\lambda$ equal to 1. We now see that the
results of our method and the Heun method are almost the same,
however, the step size of our method is larger than the one used in
Heun method. The parameters of the green line is the same as the
black line except $\lambda=10$, that is, the characteristic time is
shorter, and in this condition, the Ornstein-Uhlenbeck noise is
closer to the Gaussian noise. We can find that the \textit{resonant
noisy intensity} (the maximum of the line) shifts left when we
shorten the characteristic time. In other words, lengthening the
characteristic time can enhance the noise resistance of the system.
The blue line is the theoretical result of Eq.(41). Since the
influence of the inertia term $\ddot{x}$, we can see that the
amplitude of output $<x>$ of the theoretical result is greater than
our numerical result as shown in the green line.

\section{Conclusion}

We have proposed the bicolour rooted tree method to do the
stochastic Taylor expansion systematically. This method can be used
to solve the stochastic differential equation numerically. In this
paper, we focus on the Langevin equation which is widely used in
modern science. A high order algorithm \textit{o}(7,4.5) is derived
in this paper. Comparing with other usual algorithms, our method is
advantageous in computational efficiency and accuracy. We present
our method in the two examples. In the first example of energy
relaxation in the double well, we show our method gives the same
results as presented in other papers, and the convergence is better
than the other algorithms. In the second example, we propose an
algorithm for the time-dependent Langevin equation with the
Ornstein-Uhlenbeck noise, and the result of our algorithm is the
same as the one obtained by Heun method. It shows our algorithm is
suitable for the Langevin equation regardless of the time-dependence
of the equation. However, the readers should note that we only
provide the algorithm for Eq.(17) which satisfies the property (19)
since this property can drastically reduce the complexity of
Eq.(14). For the other type of SDEs that the property (19) can not
be held, such as the Hodgkin-Huxley model in neuroscience \cite{27},
interested readers can design their own algorithms based on the
Eq.(14). For the purpose of using our method in the more difficult
situations, one can consider the case that the diffusion
coefficients are variable. All in all, we have provided a systematic
scheme for searching the high order algorithm for the SDE and find
that it can reduce drastically when deal with the Langevin equation.

\section*{Acknowledgments}

The authors thank Prof. Yong Zhang for useful discussions and Dr.
Jigger Cheh, Shaoqiang Yu for helping with the preparation of the
paper. We also thank the anonymous referees for their helpful
advice. This research was partly supported by National Basic
Research Program of China (973 program) (Contract No. 2007CB814800)
and National Natural Science Foundation of China (Contract No.
10475067).

\end{document}